\documentclass[preprint,12pt]{elsarticle}

\usepackage{amssymb}
\usepackage{xcolor}
\usepackage{rotating}
\usepackage{graphicx}
\usepackage{multirow}

\begin{document}

\begin{frontmatter}

\title{Neutron production in ($\alpha,n$) reactions}

\author{V. A. Kudryavtsev\corref{cor1}}
\author{P. Zakhary} 
\author{B. Easeman}
\cortext[cor1]{corresponding author: V. A. Kudryavtsev, v.kudryavtsev@sheffield.ac.uk}

\address{Department of Physics and Astronomy, University of Sheffield, Sheffield, S3 7RH, United Kingdom}

\begin{abstract}
Neutrons can induce background events in underground experiments looking for rare processes. Neutrons in a MeV range are produced in radioactive decays via spontaneous fission and ($\alpha,n$) reactions, and by cosmic rays. Neutron fluxes from radioactivity dominate at large depths ($>1$~km~w.~e.). A number of computer codes are available to calculate cross-sections of ($\alpha,n$) reactions, excitation functions and neutron yields. We have used EMPIRE2.19/3.2.3 and TALYS1.9 to calculate neutron production cross-sections and branching ratios for transitions to the ground and excited states, and modified SOURCES4A to evaluate neutron yields and spectra in different materials relevant to high-sensitivity underground experiments. We report here a comparison of different models and codes with experimental data, to estimate the accuracy of these calculations.
\end{abstract}


\end{frontmatter}


\section{Introduction}
\label{intro}

Underground experiments looking for rare events, such as dark matter WIMPs, neutrinoless double-beta decay or low-energy neutrinos, are combatting various backgrounds, some of which are caused by neutrons. Neutrons may produce single-hit events in dark matter experiments indistinguishable from WIMP interactions. Neutron inelastic scattering and high-energy gammas from neutron capture give a background in a region of interest for neutrinoless double-beta decay search. Neutron interactions can also mimic signatures expected from low-energy neutrinos. 

Since these experiments are usually located at large depths underground ($>1$~km~w.~e.), we will consider here the mechanisms for neutron production relevant to underground experiments. Neutrons from atmospheric showers do not penetrate hundreds metres of rock whereas neutron fluxes from cosmic-ray muons are suppressed by several orders of magnitude relative to shallow depths. Also, muon-induced events can be tagged in experiments by a coincident detection of a muon or other particles in a muon-initiated cascade. This makes radioactive decays the main source of 
neutron background at large depths. Two processes contribute to neutron production of this origin: spontaneous fission and ($\alpha,n$) reactions. 

Spontaneous fission (SF), as described by Watt's formulae \cite{watt}, gives the same neutron yield for all materials and depends only on the concentration of the fissioning isotope. Among naturally occurring radioactive isotopes, only fission of $^{238}$U contributes significantly to neutron production. Although the probability of SF of $^{238}$U is about $5\times10^{-7}$ compared to alpha-decay rate, the neutron yield from this process dominates over that from ($\alpha,n$) reactions for high-$Z$ materials where the neutron production is highly suppressed due to the Coulomb barrier. In practice neutron events from the SF process can be tagged in a detector or a veto system due to simultaneous emission of several neutrons and gamma-rays. 

Neutron yield from ($\alpha,n$) reactions depends on the alpha energy, cross-section of the reaction and alpha energy loss in a particular material. Two radioactive decay chains are critical in the calculation of neutron background for underground experiments. These chains start with the parent isotopes of $^{238}$U and $^{232}$Th and consist of 6 and 8 alpha-decays, respectively. The energies of all alphas in the decays need to be considered. The secular equilibrium in the decay chains is not always in place. The decay chain of $^{235}$U also contributes to the neutron production but this contribution is relatively small due to a small abundance of $^{235}$U in natural uranium (0.72\%).

The cross-sections of ($\alpha,n$) reactions are isotope-dependent and can be calculated using nuclear physics codes, such as EMPIRE \cite{empire} or TALYS \cite{talys}, or taken from experimental data where available. The energy threshold of these reactions is determined by the $Q$-value of the reaction and the Coulomb barrier that suppresses the reaction probability even if the alpha energy is above the threshold determined by the $Q$-value. Hence the ($\alpha,n$) reactions are important for low- and medium-$Z$ nuclei while the neutron yield per unit concentration of a radioactive isotope from elements heavier than copper is quite small. 

Alpha-particles produced in radioactive decays are quickly losing energy in a material via ionisation and excitation of atoms, thus reducing the probability of neutron production. Energy loss of alphas is well understood and is taken into account in all codes dealing with neutron yield calculation.

Specialist computer codes have been developed over a number of years to calculate neutron yields and energy spectra from ($\alpha,n$) reactions. The nuclear physics code SOURCES4A/4C \cite{sources4} has been used for a long time in a number of applications. The code is based on libraries that contain alpha lines from most radioactive isotopes, energy losses of alphas in different materials and cross-sections of ($\alpha,n$) reactions. A big advantage of the code is that it is flexible and allows the user to choose a cross-section of ($\alpha,n$) reaction from different sources, and more cross-sections can also be added to the library. Due to the recent needs of calculating neutron background for various underground experiments, new codes have been written and used by some experiments (see for instance, \cite{usd,neucbot}). Some of them use cross-sections from TALYS and show results quite different from SOURCES4 (for more detailed comparison see \cite{scorza,neucbot,fernandes,vk-idm2018}). There is also a recent adaptation of GEANT4 to handle cross-sections and neutron energy spectra files from JENDL  and TENDL libraries in ENDF format \cite{mendoza2019}.

In this paper, we present a detailed study of neutron production as calculated by SOURCES4A with different cross-sections that are also compared to available experimental data. A brief description of SOURCES4A code and modifications to the code is given in Section \ref{sec-sources4} together with the description of the cross-sections used. This Section also includes comparison of different cross-sections used in SOURCES4A with experimental data. Results of neutron yield and neutron spectra calculations are shown in Section \ref{sec-n-comp} and the conclusions are given in Section \ref{sec-conclusions}.

\section{SOURCES4A and cross-sections for ($\alpha,n$) reactions}
\label{sec-sources4}

The computer code SOURCES4A \cite{sources4a} uses the libraries of alpha emission lines from radioactive isotopes, cross-sections of ($\alpha,n$) reactions either from calculations or experimental data, excitation functions (probabilities of transitions into excited states) and energy loss of alphas in different materials, to calculate the neutron production rate and energy spectra of emitted neutrons. The most recent version is SOURCES4C \cite{sources4} but for historical reasons we are using the older version SOURCES4A \cite{sources4}. The release notes provided by the authors and the tests done previously showed that, provided the same cross-sections and excitation functions are used, there is no difference between the results from the two versions for our simulations. 

The original code calculates neutron production for alpha energies below 6.5 MeV and is not suitable for our purpose since alphas from radioactive decays have energies up to about 9 MeV. The code SOURCES4A has been modified to extend the alpha energy range to about 10 MeV \cite{tomasello2008}. 'New' cross-sections and excitation functions calculated using the EMPIRE2.19 code \cite{empire} have been added to the code library covering the range of alpha energies up to 10 MeV \cite{carson,tomasello2008,lemrani,tomasello2010,tomasello-thesis}. All these changes have been made for the version SOURCES4A and, since no changes were observed with the newer version of the code, we continue to use SOURCES4A for our calculations. A comparison of cross-sections from EMPIRE2.19 with experimental data was published in Refs. \cite{tomasello2008,tomasello-thesis} and the results of neutron yield calculations with modified SOURCES4A were used for a number of dark matter experiments (see, for example, Refs. \cite{eureca,edelweiss,lz,xenon1t}). The accuracy of the calculation was estimated to be about 20\% based on the comparison of neutron yields obtained with different sets of cross-sections \cite{tomasello2008}. 

The user input to SOURCES4A includes material composition (where the alpha sources are located), isotopic composition for each element (only isotopes with cross-sections present in the code library can be included) and either the energy of the alpha-particle or the radioactive isotope (or several isotopes in the case of decay chains, for instance) with the number of atoms in a sample. For application in low-background experiments an option of the thick target neutron yield was chosen, meaning that the size of the material sample is much bigger than the range of alphas and edge effects can be neglected. 

The output of SOURCES4A includes several files that return the neutron yield and spectra for the sum of the ground and all excited states, as well as neutron spectra for individual states. In the case of decay chains, neutron production from individual radioactive isotopes on each isotope in the material sample is also calculated. SOURCES4A/4C do not calculate gamma production but the total energy transferred to gammas can be calculated from the energy of the excited states if required.

More recently new nuclear physics codes have become available for calculating ($\alpha,n$) reaction cross-sections, such as TALYS \cite{talys} and newer versions of EMPIRE \cite{empire} and below we present the comparison of previous neutron yield calculations (with cross-sections from EMPIRE2.19) to those with newer sets of cross-sections (from TALYS1.9 and EMPIRE3.2.3). Apart from the open source code the authors of TALYS also provide calculated cross-sections and various energy-angle distributions as TENDL libraries that can be used for various applications. Unfortunately, the data in TENDL are different from those required for SOURCES4A input libraries. In this work the TALYS code itself has been used to calculate cross-sections and transition probabilities that were then used as input to SOURCES4A.

All cross-sections and transition probabilities were calculated with a 0.1~MeV step in alpha energy.

To validate the models for calculating cross-sections of ($\alpha,n$) reactions we have compared these cross-sections with experimental data where available. The model for calculating cross-sections in EMPIRE2.19 has been described in Ref. \cite{tomasello2008}. The description of the TALYS code can be found in Ref. \cite{talys} and references therein. We have used the default input parameters in TALYS for calculation of ($\alpha,n$) cross-sections and branching ratio of transitions to different states, effectively leaving to the code and its developers to choose the nuclear model and its parameters. For the most recent version of EMPIRE3.2.3 \cite{empire3.2,empire3.2.3} several models have been tested. 

The EMPIRE code accesses the Reference Input Parameter Library (RIPL)~\cite{capote2009}, maintained by the International Atomic Energy Agency (IAEA), and extract nuclear masses, level densities, Optical Model Potentials (OMP), decay schemes, experimental data, etc. The Fermi Gas Model has been chosen to calculate the nuclear level density. EMPIRE handles three major nuclear reaction mechanisms, namely Direct Reaction (DR), Pre-equilibrium Emission (PE) and Compound Nucleus (CN) decay, the sum of which gives the total reaction cross-section.

\begin{itemize}
    \item In the DR mechanism, the code uses the spherical OMP with an option to choose a particular model that is best suited for a specific projectile particle, nuclear reaction, atomic weight of a target nucleus and energy range. Most of the target isotopes are covered by the widely adopted McFadden-Satchler (MS) potential \cite{mcfadden1966,simon2017}, which has been used as a benchmark also for TALYS, especially for $\alpha$-induced reactions with incident energy up to 10 MeV. It describes an interaction between an $\alpha$ particle and a target nucleus by a complex mean-field potential. Subsequently, it divides the reaction into an elastic and inelastic scattering channels.
    
    \item In the PE mechanism, the exciton model \cite{griffin1966,ribansky1973} is used with its mean free path multiplier set to handle different projectiles. The code then calculates emission spectra and populate the discrete levels of all target isotopes. The contribution of this mechanism increases with incident energy.
    
    \item In the CN mechanism, an advanced implementation of Hauser-Feshbach (HF) theory~\cite{hauser1952} is used. It calculates the decay of the compound nucleus into discrete levels of residual nucleus with the emission of light nuclei, nucleons and other particles.

\end{itemize}

EMPIRE cross-sections depend primarily on the chosen OMP and several models have been tested.

\begin{figure}[h!]
  \begin{minipage}{8cm}
\includegraphics[width=8cm]{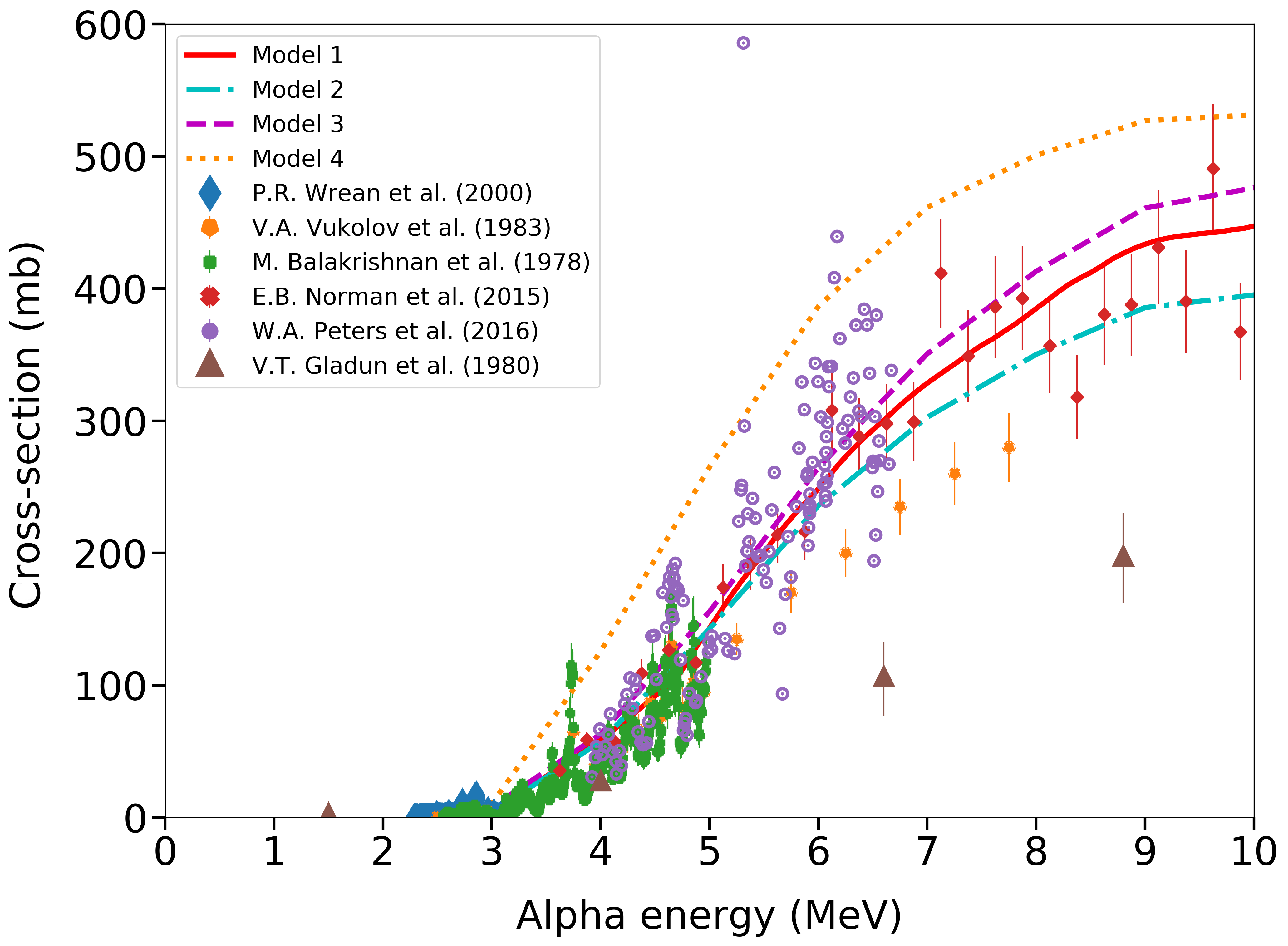}
\caption{\label{fig:cs-models-fluorine}
Cross-sections of ($\alpha,n$) reactions in fluorine ($^{19}$F) as calculated in different models in EMPIRE3.2.3. Model 1 -- MS-OMP with DR contribution, Model 2 -- MS-OMP without DR contribution, Model 3 and Model 4 employ OMPs from Refs. \cite{huizenga,avrigeanu} without and with the DR contribution, respectively. The data for fluorine are taken from Wrean et al. \cite{wrean2000}, Vukolov et al. \cite{vukolov1983}, Balakrishnan et al. \cite{balakrishnan1978}, Norman et al. \cite{norman2015}, Peters et al. \cite{peters2016} and Gladun et al. \cite{gladun1980}. (Note that Ref.~\cite{norman2015} has reported previous measurements of thick target neutron yield used to evaluate the cross-section of the ($\alpha,n$) reaction on fluorine \cite{norman1984} shown here.)
}
\end{minipage}\hspace{0.5cm}%
  \begin{minipage}{8cm}
  \includegraphics[width=8cm]{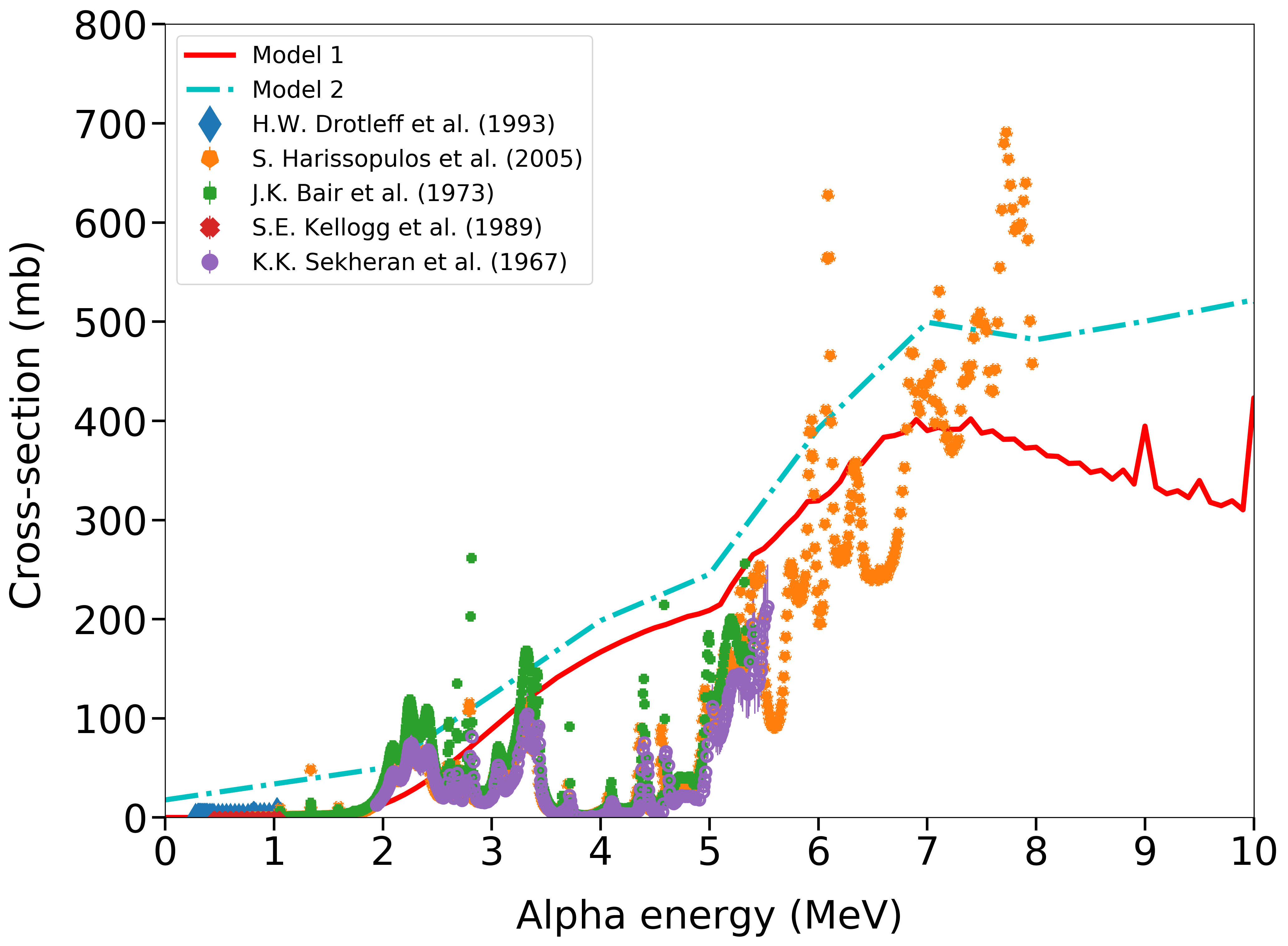}
\caption{\label{fig:cs-models-carbon}
Cross-sections of ($\alpha,n$) reactions in $^{13}$C as calculated in different models in EMPIRE3.2.3. Model 1 -- OMP from \cite{wilmore} without DR contribution, Model 2 -- OMP from \cite{wilmore} with DR contribution. The data for $^{13}$C are from Drotleff et al. \cite{drotleff1993}, Harissopulos et al. \cite{harissopulos2005}, Bair et al. \cite{bair1973}, Kellogg et al. \cite{kellogg1989} and Shekharan et al. \cite{shekharan1967}. (Note that a correction to the original results of \cite{harissopulos2005} has been applied in Ref. \cite{mohr2018} leading to a smaller cross-section, not shown here.)
}
  \end{minipage}
\end{figure}

Figures \ref{fig:cs-models-fluorine} and \ref{fig:cs-models-carbon} show the comparison between different EMPIRE models and experimental data for $^{19}$F (the only stable and hence, naturally occurring isotope of fluorine) and $^{13}$C. Fluorine ($^{19}$F) has been selected as being known to give high neutron yield. Carbon has two naturally occurring stable isotopes but $^{12}$C has a high threshold for $(\alpha, n)$ reactions and does not contribute to the neutron yield if alphas are produced in radioactive decays of U/Th and their progeny. $^{13}$C has a low threshold and, despite its low abundance, may contribute significantly to the neutron production in carbon-abundant materials, such as scintillators, acrylic, polyethylene and many plastics.

Figure \ref{fig:cs-models-fluorine} shows the ($\alpha,n$) reaction cross-section for $^{19}$F using the global alpha-potentials in the MS models and in models described in Refs. \cite{huizenga,avrigeanu}.  Model 1 and Model 2 use MS-OMP with and without the DR contribution, respectively. Similarly, Model 3 and Model 4 employ OMPs from Refs. \cite{huizenga,avrigeanu} without and with the DR contribution, respectively. Clearly, Model 4 overestimate the cross-section. MS-OMP with DR contribution (Model 1) gives reliable results and, in fact has been recommended by the code developers as the best suited for a number of isotopes including $^{19}$F. In future calculations we will use a model that was recommended by the code developers and call it Model 1 unless explicitly stated otherwise.

Figure \ref{fig:cs-models-carbon} shows the ($\alpha,n$) reaction cross-section for $^{13}$C as calculated with the
OMP from Ref. \cite{wilmore} without (Model 1) and with (Model 2) the DR contribution, respectively. These models are meant to be used for neutron-induced reactions. Yet, they have been used here since the MS-OMP does not cover reactions on light elements ($A<16$). Discrepancies might arise, whenever a neutron OMP is used for $\alpha$-induced reactions, because of the missing Coulomb field term, which is absent for neutron scattering. In addition, uncertainties in the imaginary part of the potential could be substantially dependent on the projectile properties \cite{wilmore}. It is worth mentioning that sometimes models based on neutron-induced reactions at higher energies can still provide a better agreement with experimental data compared to models with OMPs specifically targeting $\alpha$-induced reactions.

The differences between various sets of experimental data do not always allow us to choose the best nuclear physics model based on such a comparison so, in the calculations described below, we have used the models recommended by the authors of EMPIRE \cite{empire} (identified as Model 1 in Figures \ref{fig:cs-models-fluorine} and \ref{fig:cs-models-carbon}). Note that the optimum models (chosen OMP) are different for different isotopes as they were tuned to specific nuclei. 

\begin{figure}[h!]
  \begin{minipage}{8cm}
\includegraphics[width=8cm]{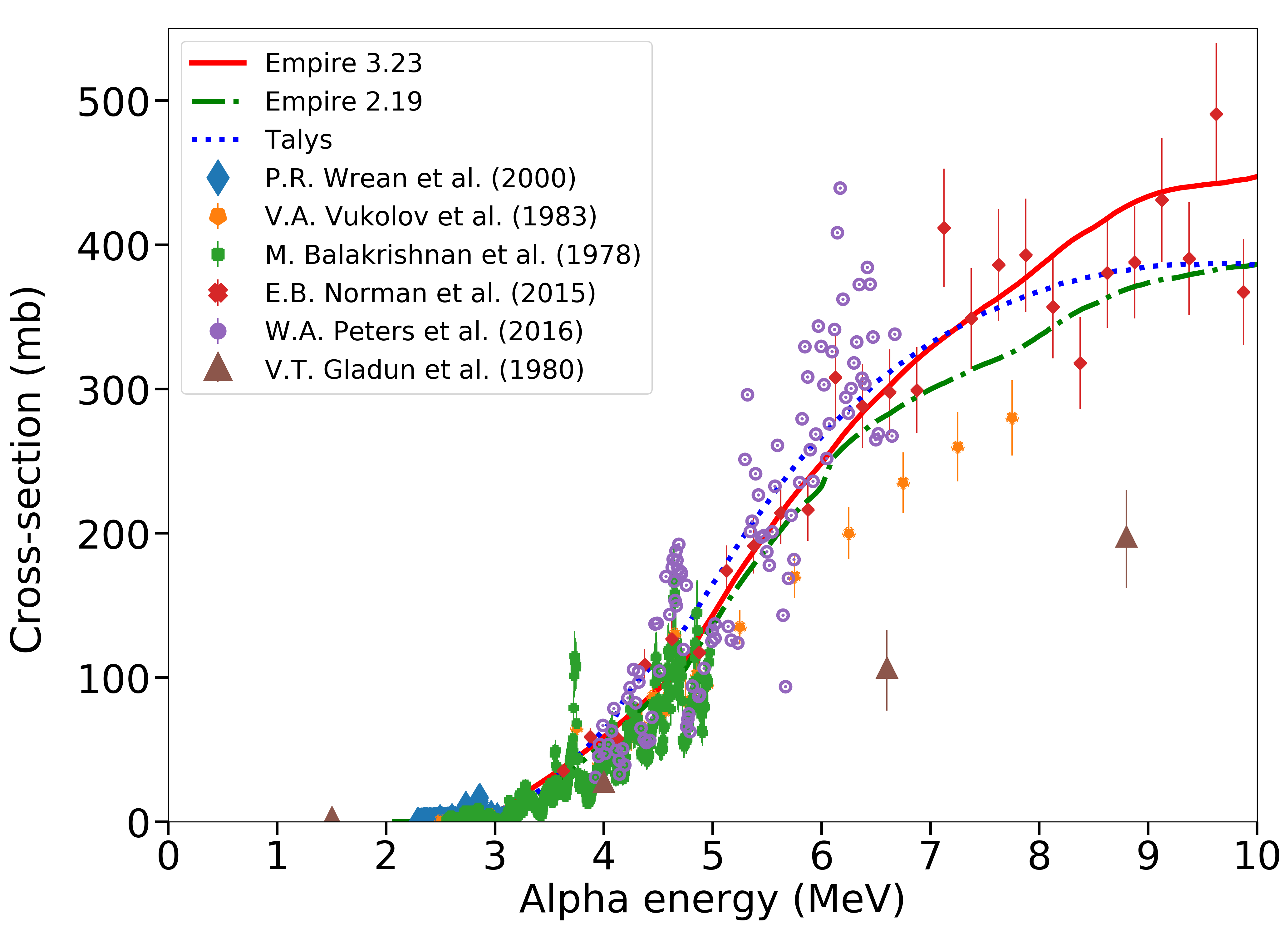}
  \caption{  \label{fig:cs-codes-fluorine} 
Cross-sections of ($\alpha,n$) reactions for fluorine ($^{19}$F) as calculated by EMPIRE2.19, EMPIRE3.2.3 and TALYS1.9 in comparison with experimental data. The data for fluorine are taken from Wrean et al. \cite{wrean2000}, Vukolov et al. \cite{vukolov1983}, Balakrishnan et al. \cite{balakrishnan1978}, Norman et al. \cite{norman2015}, Peters et al. \cite{peters2016} and Gladun et al. \cite{gladun1980}.. 
}
\end{minipage}\hspace{0.5cm}%
  \begin{minipage}{8cm}
  \includegraphics[width=8cm]{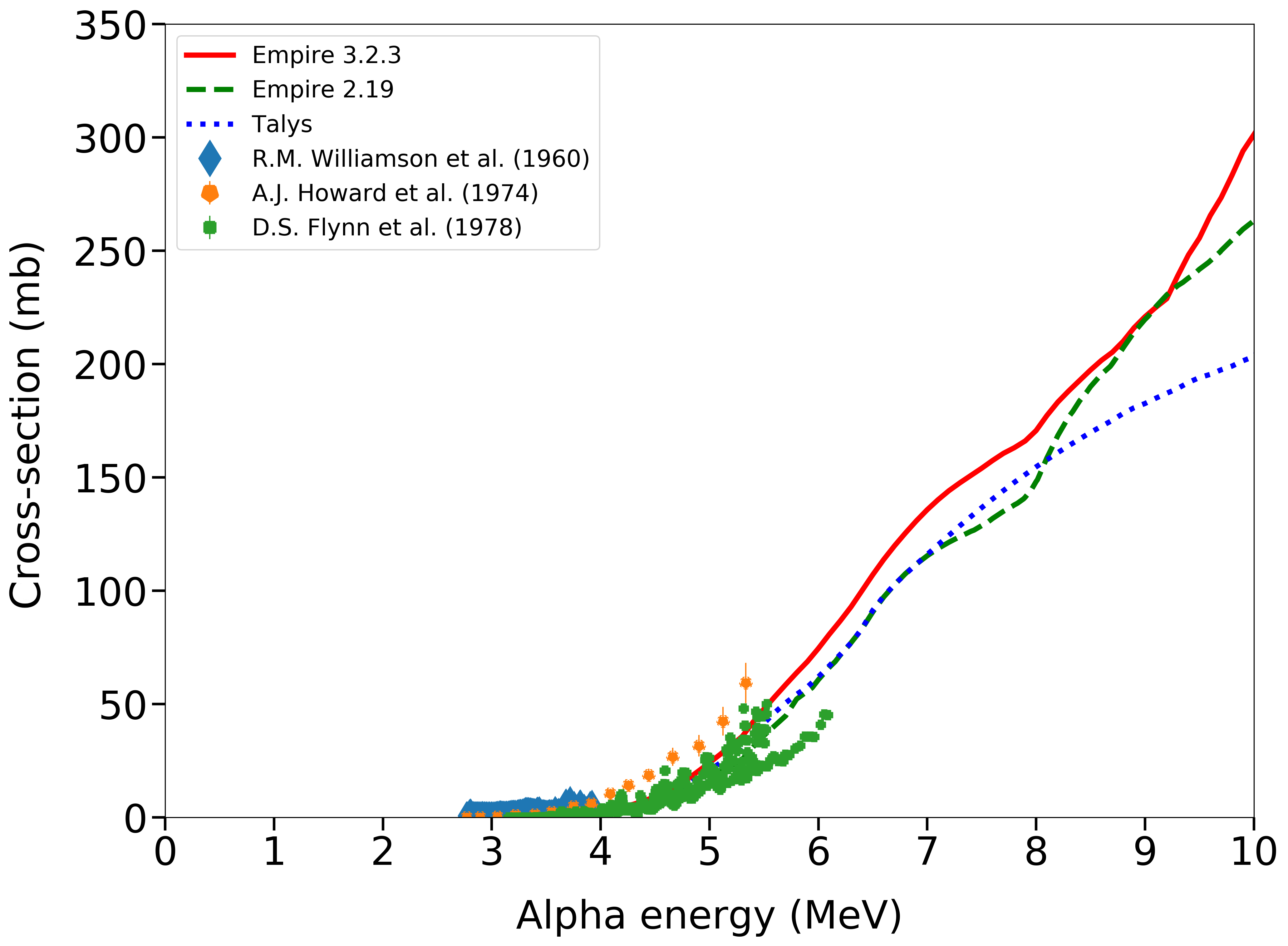}
\caption{  \label{fig:cs-codes-aluminium} 
Cross-sections of ($\alpha,n$) reactions for aluminium ($^{27}$Al) as calculated by EMPIRE2.19, EMPIRE3.2.3 and TALYS1.9 in comparison with experimental data. The data for aluminium are taken from Williamson et al. \cite{williamson1960}, Howard et al. \cite{howard1974} and Flynn et al. \cite{flynn1978}.}
  \end{minipage}
\end{figure}

Figures~\ref{fig:cs-codes-fluorine} and \ref{fig:cs-codes-aluminium} show the comparison of the cross-sections as calculated by EMPIRE2.19, EMPIRE3.2.3 and TALYS1.9 with experimental data. Left plot shows the results for fluorine and the right plot shows the results for aluminium (namely, for $^{27}$Al, as the only stable and naturally occurring isotope of aluminium). Cross-sections calculated with a newer version of EMPIRE3.2.3 are bigger than those with the older version of EMPIRE2.19. This statement is true for almost all isotopes tested. Similarly, TALYS cross-sections are bigger than those from EMPIRE2.19 for most isotopes tested (an exception is aluminium, shown in Figure~\ref{fig:cs-codes-aluminium}, for which TALYS and EMPIRE2.19 give similar results up to about 8~MeV). EMPIRE3.2.3 gives higher cross-sections than TALYS for aluminium but smaller for fluorine for most energies in the range of interest for alphas from radioactive decays. There is no specific tendency when comparing EMPIRE3.2.3 and TALYS for other isotopes. Despite several sets of data collected for fluorine, there is no obvious choice of the model/code that would match all these data sets. The difference between some of the data sets exceeds the variations between the models/codes. This conclusion holds for most isotopes studied.

\begin{figure}[h]
  \begin{minipage}{8cm}
\includegraphics[width=8cm]{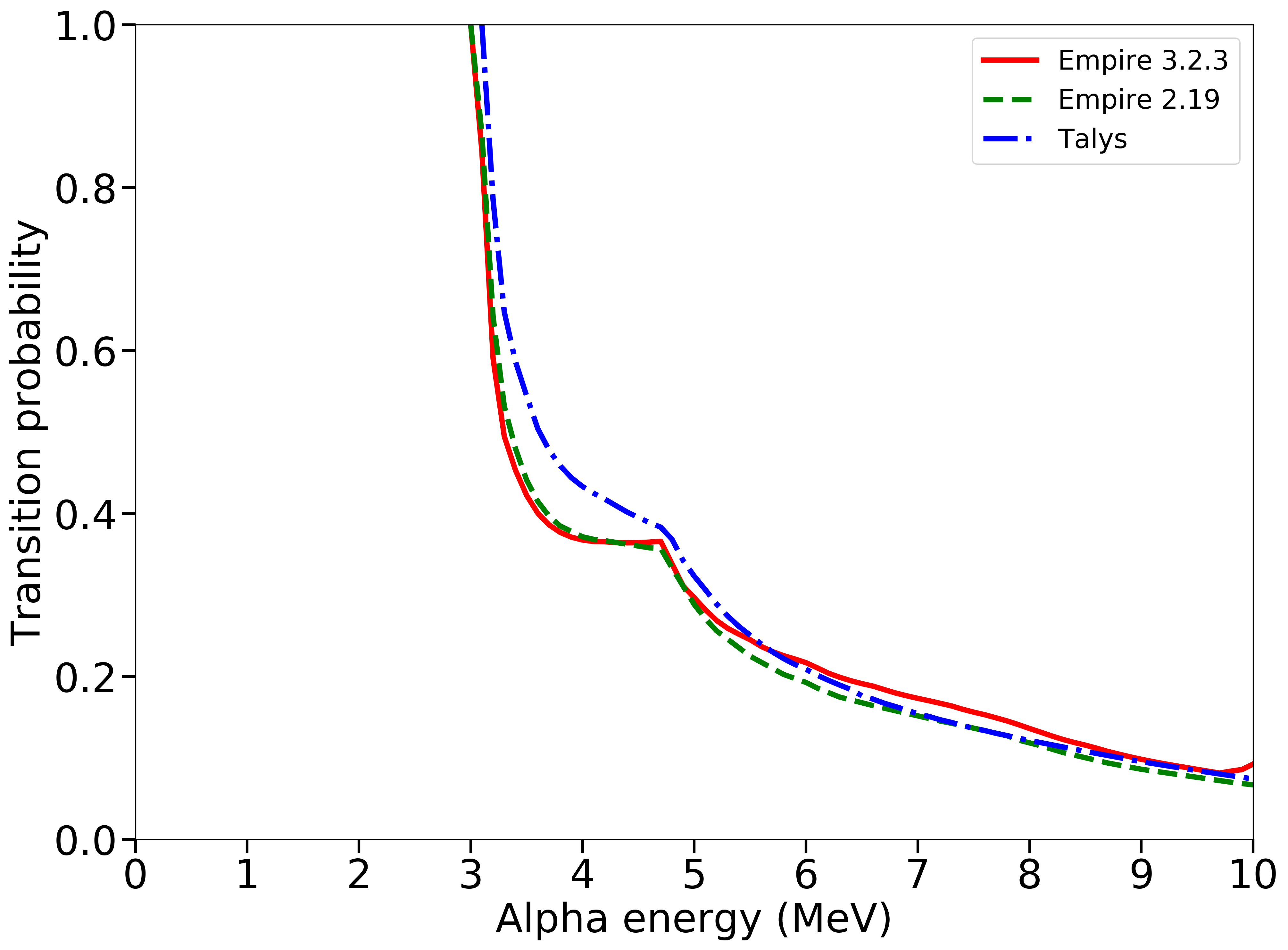}
  \caption{  \label{fig:excitations-ground} 
Transition probability as a function of alpha energy for the ground state of $^{22}$Na for the ($\alpha,n$) reaction on $^{19}$F as calculated by EMPIRE2.19, EMPIRE3.2.3 and TALYS1.9.}
\end{minipage}\hspace{0.5cm}%
  \begin{minipage}{8cm}
  \includegraphics[width=8cm]{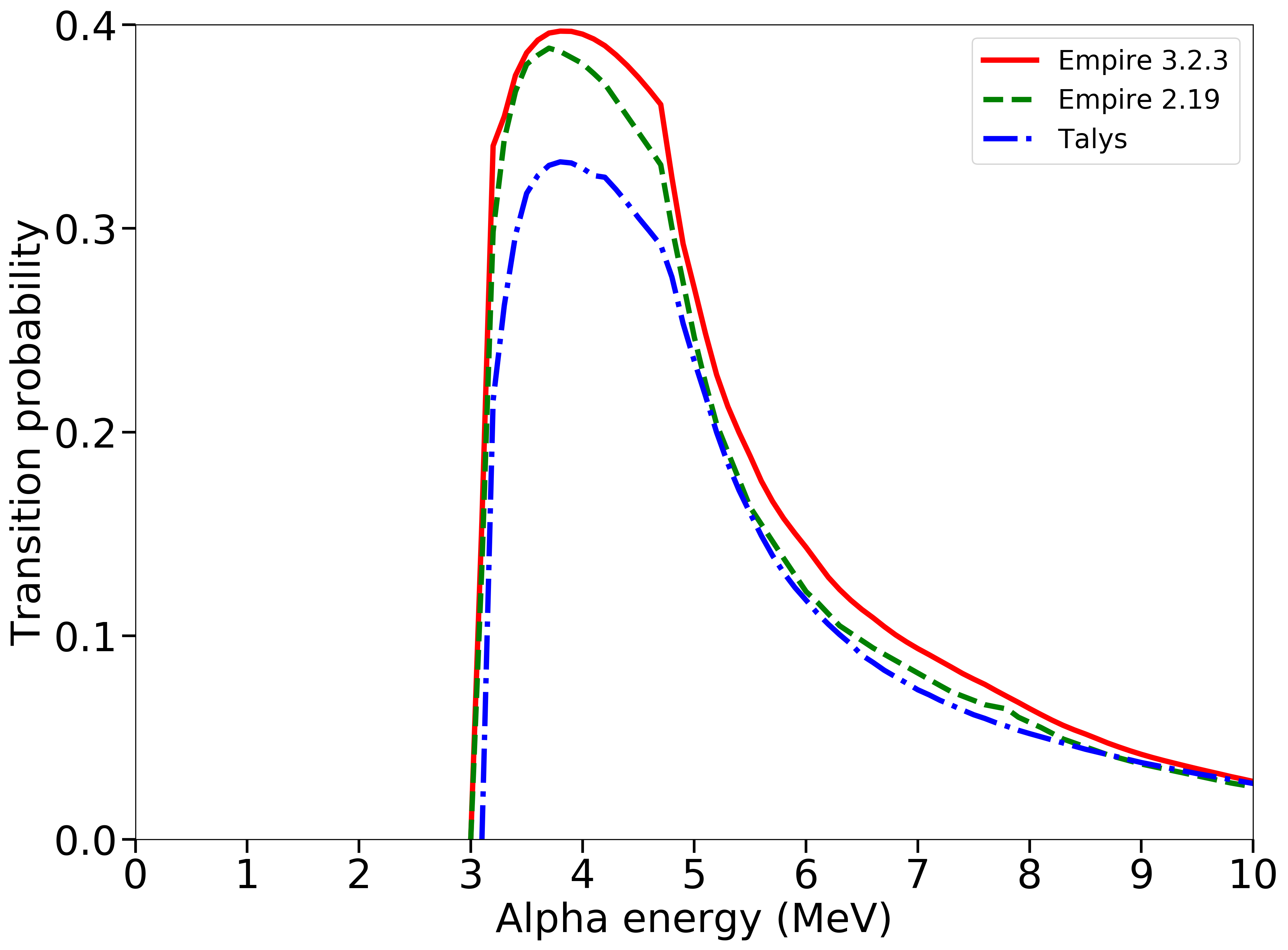}
  \caption{  \label{fig:excitations-excited} 
Transition probability as a function of alpha energy for the first excited state of $^{22}$Na for the ($\alpha,n$) reaction on $^{19}$F as calculated by EMPIRE2.19, EMPIRE3.2.3 and TALYS1.9.}
  \end{minipage}
\end{figure}

Neutron energy spectra from ($\alpha,n$) reactions depend also on the excitation functions or probabilities of transition of a nucleus to excited states followed by the emission of gamma-rays. If the final state nucleus is left in an excited state, less energy is transferred to a neutron. SOURCES4A code does not calculate the gamma-ray production but neutron spectra are calculated for individual ground and excited states and for the sum of the above. Figures \ref{fig:excitations-ground} and \ref{fig:excitations-excited} show transition probabilities for fluorine to the ground and the 1st excited state for 3 different codes. The 3 models show a similar behaviour for transition probabilities. This conclusion is also true for higher excited states. In future we will present only the neutron spectra for the sum of all states.

\section{Results and discussion}
\label{sec-n-comp}

\begin{figure}[h]
  \begin{minipage}{8cm}
\includegraphics[width=8cm]{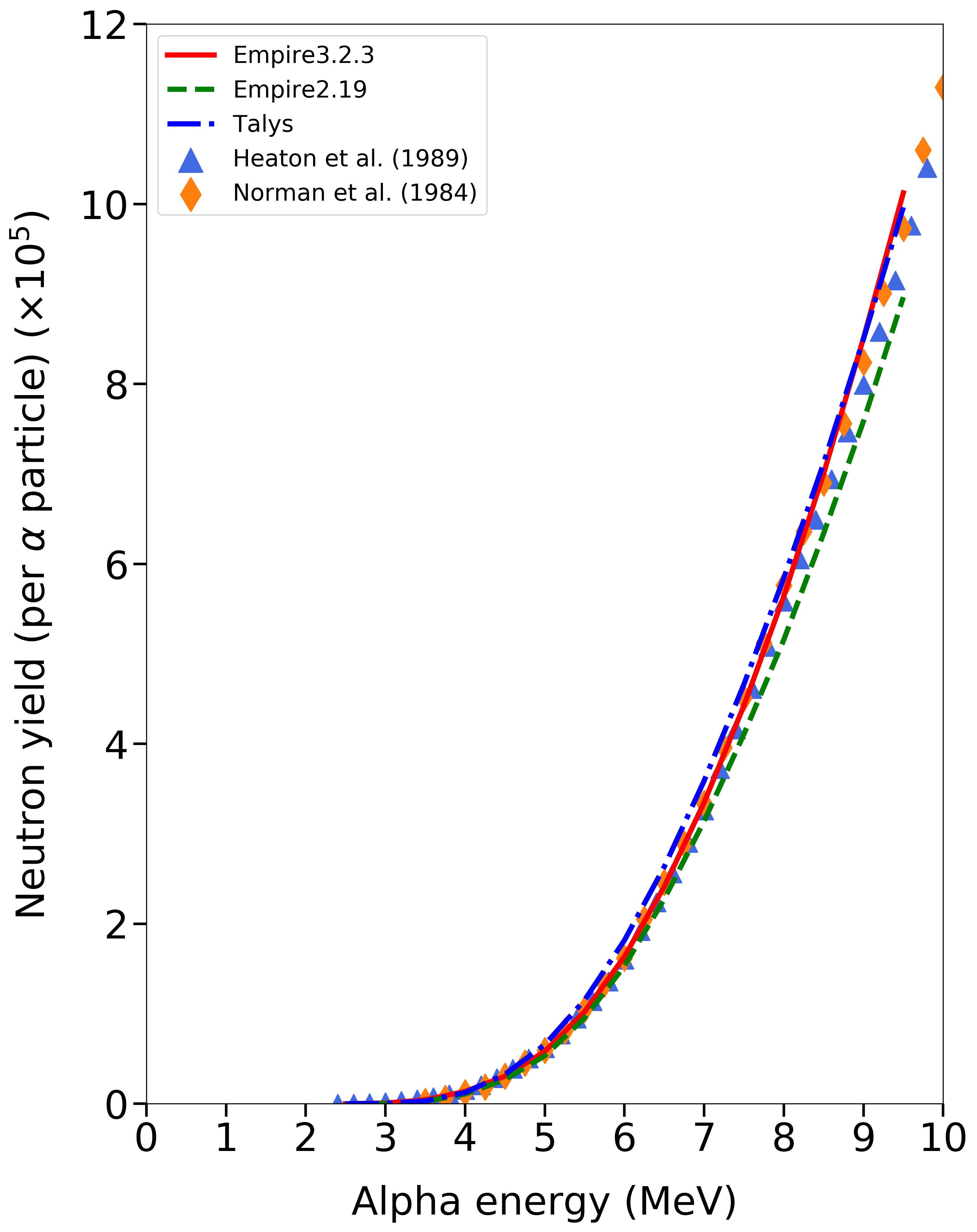}
  \caption{  \label{fig:nyield-beam-fluorine} 
Neutron yield from ($\alpha,n$) reactions in fluorine as a function of alpha energy. Data are taken from Heaton et al. \cite{heaton1988} and Norman et al. \cite{norman1984}.
}
\end{minipage}\hspace{0.5cm}%
  \begin{minipage}{8cm}
  \includegraphics[width=8cm]{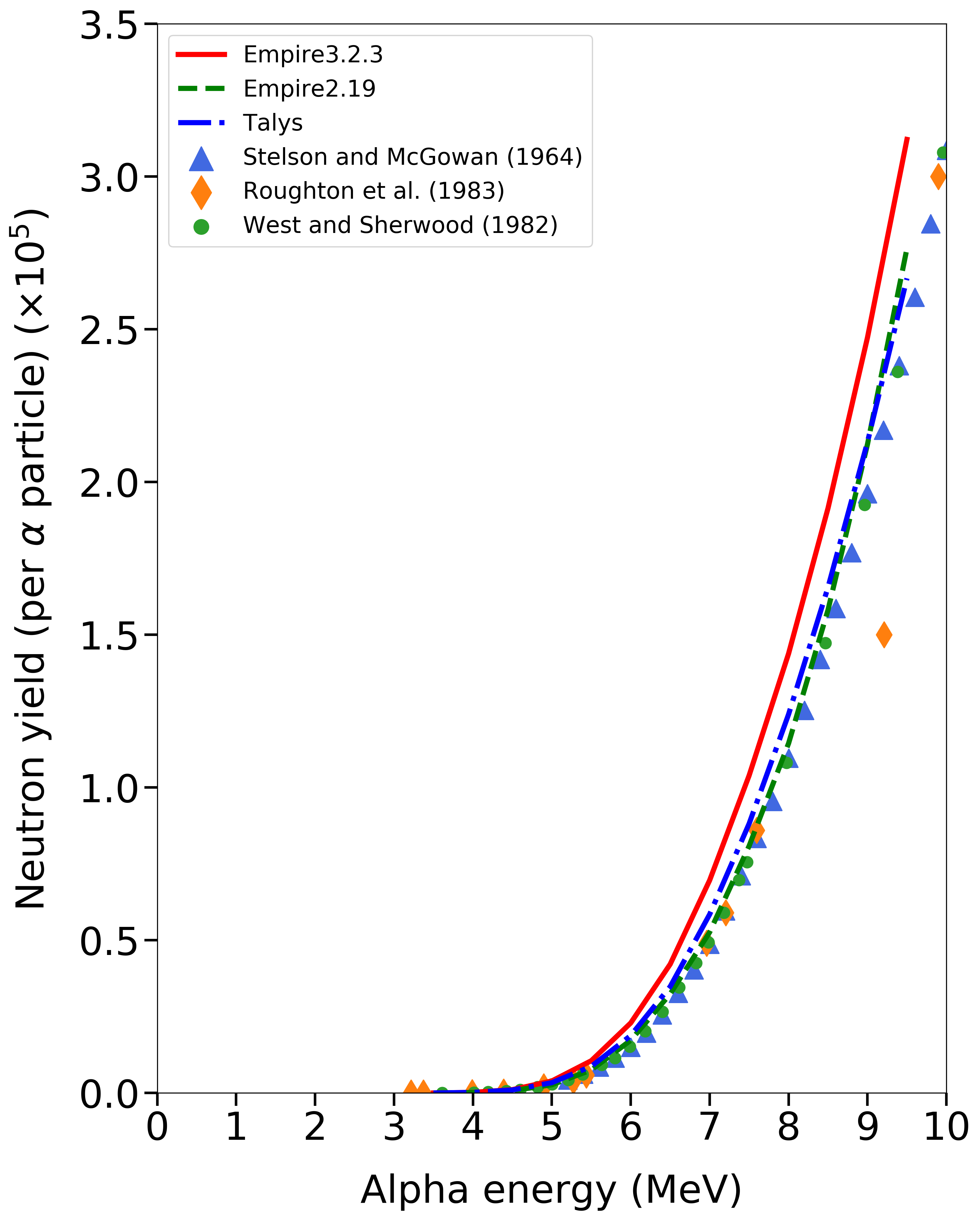}
  \caption{  \label{fig:nyield-beam-aluminium} 
Neutron yield from ($\alpha,n$) reactions in aluminium as a function of alpha energy. Data are taken from Stelson and McGowan \cite{stelson1964}, Roughton et al. \cite{roughton1983}, and West and Sherwood \cite{west1982}.}
  \end{minipage}
\end{figure}

SOURCES4A returns total neutron spectra originated from ($\alpha,n$) reactions from the decay of radioactive isotopes and also from ($\alpha,n$) reactions occurring on individual isotopes present in a material specified by the user. Firstly we compare neutron yields from alphas with fixed energies originated from different materials. Figure \ref{fig:nyield-beam-fluorine} and \ref{fig:nyield-beam-aluminium} show neutron yields in fluorine and aluminium, respectively, as a function of the initial alpha energy. All three codes, EMPIRE3.2.3, EMPIRE2.19 and TALYS1.9 agree reasonably well with the data for fluorine with the latest version EMPIRE3.2.3 and TALYS1.9 giving slightly higher neutron yields than data and the previous version of EMPIRE2.19. For aluminium, TALYS1.9 and EMPIRE2.19 agree better with data than EMPIRE3.2.3.

In underground experiments, neutron background originates from the decay chains of uranium and thorium. Below we present the results of calculations of neutron yields and energy spectra from the whole chains assumed to be in secular equilibrium, as well as separate calculations for early and late uranium sub-chains. The split of $^{238}$U chain into early and late is very likely happening at $^{226}$Ra which decay is the first one in the late sub-chain. All isotopes in decay chain of $^{235}$U (except $^{235}$U itself) have very short half-lives compared to $^{235}$U so this chain is not split and is added to the early sub-chain of $^{238}$U and to the whole $^{238}$U chain. Although $^{235}$U abundance is only 0.72\% in natural uranium, its chain contains alphas of high energies (up to about 8~MeV) that can make a non-negligible contribution to the early sub-chain. In most cases, however, SF of $^{238}$U will give a significantly higher contribution to the early sub-chain. $^{232}$Th is assumed to be in equilibrium. 

\begin{figure}[h]
  \begin{minipage}{8cm}
\includegraphics[width=8cm]{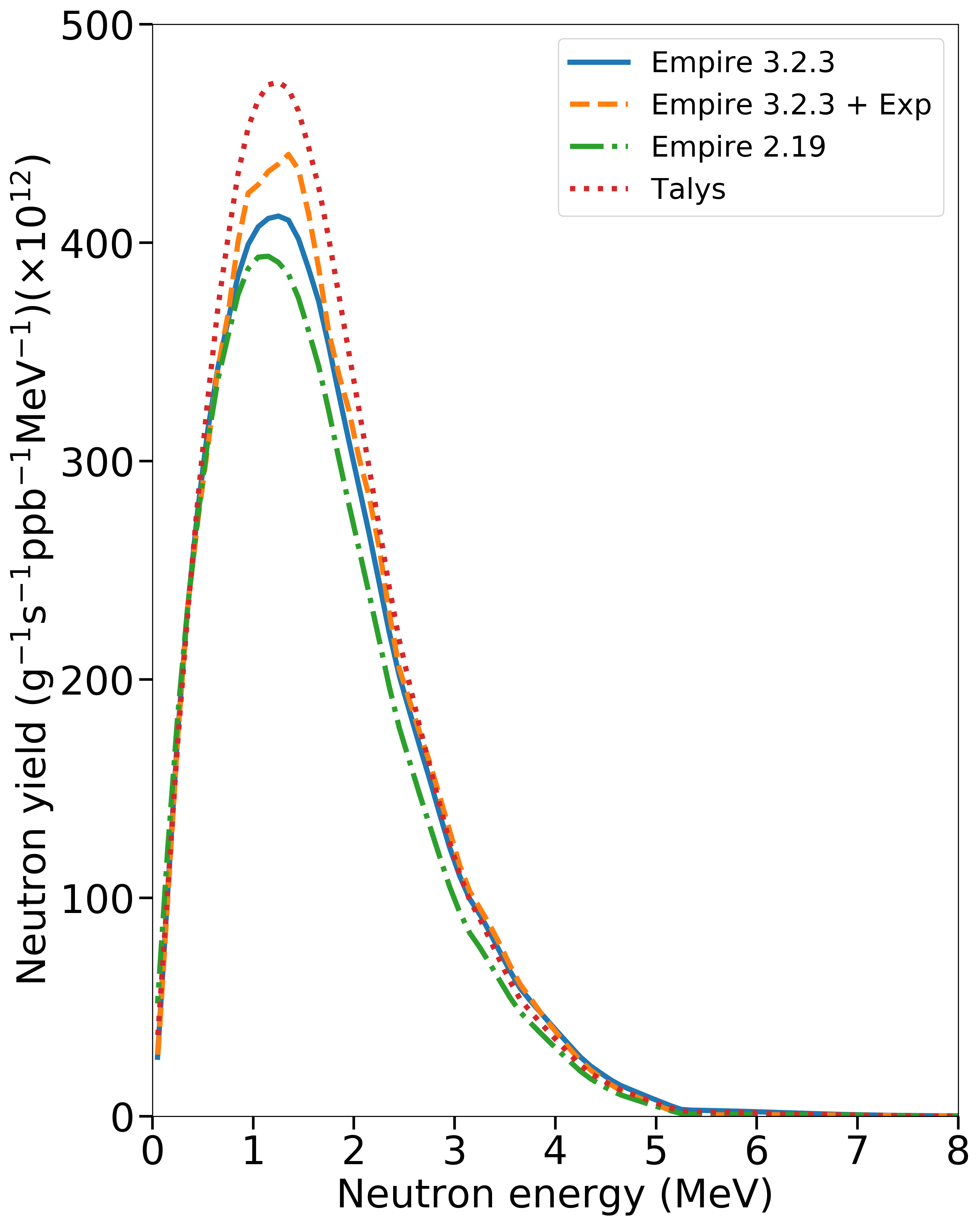}
  \caption{  \label{fig:nsp-ptfe} 
Neutron spectra from ($\alpha,n$) reactions in PTFE as calculated by EMPIRE2.19, EMPIRE3.2.3 and TALYS1.9 from uranium decay chain ($^{238}$U and $^{235}$U) in equilibrium.}
\end{minipage}\hspace{0.5cm}%
  \begin{minipage}{8cm}
  \includegraphics[width=8cm]{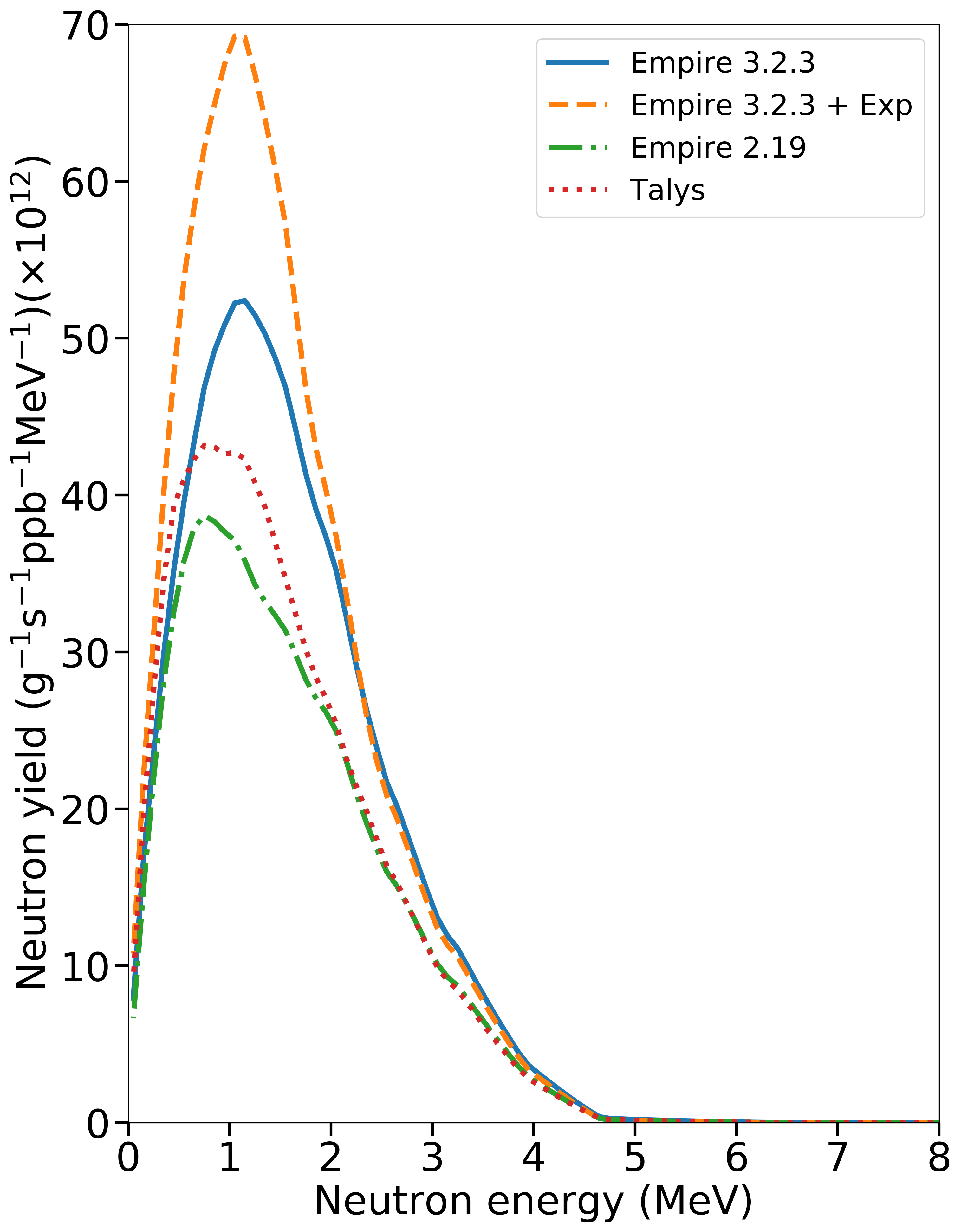}
  \caption{  \label{fig:nsp-ceramics} 
Neutron spectra from ($\alpha,n$) reactions in aluminium oxide as calculated by EMPIRE2.19, EMPIRE3.2.3 and TALYS1.9 from uranium decay chain ($^{238}$U and $^{235}$U) in equilibrium.}
  \end{minipage}
\end{figure}

Figure~\ref{fig:nsp-ptfe} and \ref{fig:nsp-ceramics} show the neutron spectra produced in the decay of the whole chain of uranium ($^{238}$U and $^{235}$U) in PTFE (C$_2$F$_4$) and aluminium oxide (ceramics, Al$_2$O$_3$), respectively. For PTFE, almost all contribution comes from fluorine and carbon affects only alpha transport. As the threshold for neutron production on $^{12}$C exceeds energies of alphas produced in radioactive decays, only $^{13}$C gives a small contribution to the neutron yield from PTFE. For ceramics, the majority of neutrons come from aluminium while small abundances of $^{17}$O and $^{18}$O give small fraction of neutrons visible at high energies (above 4 MeV). The curves labelled `EMPIRE3.2.3 + Exp' shows the neutron spectra obtained with a combination of cross-sections from experimental data (\cite{peters2016} for PTFE and \cite{howard1974} for ceramics) and EMPIRE3.2.3 calculations. The experimental data are used where available, replaced with EMPIRE3.2.3 calculations at higher energies. A higher neutron yield for a curve marked as `EMPIRE3.2.3 + Exp' in Figure~\ref{fig:nsp-ceramics} is due to high experimental cross-section on aluminium as shown in Figure~\ref{fig:cs-codes-aluminium}. If another set of measured cross-section data is used, the neutron yield and spectrum will be different, for instance a cross-section from Ref.~\cite{flynn1978} gives a lower neutron yield and spectrum.

The results of calculations from all three codes are presented in Table \ref{table:results}. Rows 'EMPIRE3.2.3 + Exp' include results obtained using cross-sections from measurements where available, replaced by EMPIRE3.2.3 calculations at higher energies (see Table \ref{table:results} for details).

\begin{table}
\begin{center}
\caption{Neutron yield from ($\alpha,n$) reactions in different materials. The column ``Formula / composition" gives the composition of the material sample to calculate neutron spectra, either with a chemical formula or with an abundance of elements (in \% by the number of atoms, not mass) given in brackets. Only elements with the abundance greater than 1\% are shown in the Table (with the accuracy of 1\%) but more accurate composition was used in the calculations. Neutron yield (columns 4--7) is shown as the number of neutrons per gram of material per second per ppb of U or Th concentration. Uranium and thorium decay chains are assumed to be in equilibrium in columns 4 and 7. In columns 5 and 6 early and late uranium sub-chains are shown separately. $^{235}$U is added to the early sub-chain and the whole chain of uranium. Rows 'EMPIRE3.2.3 + Exp' include results obtained using cross-sections from measurements where available, replaced by EMPIRE3.2.3 calculations at higher energies. Given a large spread of experimental data for cross-sections, only one measurement was used for each isotope and these measurements were taken from: \cite{peters2016} for $^{19}$F, \cite{harissopulos2005} for $^{13}$C, \cite{gruhle1972} for $^{14}$N, \cite{howard1974} for $^{27}$Al, \cite{vukolov1983a} for $^{17}$O and $^{18}$O, \cite{cheng1980} for $^{28}$Si, \cite{flynn1978} $^{29}$Si and $^{30}$Si, \cite{morton1994} for $^{50}$Cr, \cite{tims1991} for $^{54}$Fe, \cite{tims1993} for $^{55}$Mn, \cite{stelson1964} for $^{60}$Ni and $^{62}$Ni, \cite{zyskind1979} for $^{64}$Ni, \cite{vlieks1974} for $^{46}$Ti and \cite{baglin2004} for $^{48}$Ti. Spontaneous fission is significant for $^{238}$U only and is independent of the material with a neutron yield of $1.353 \times 10^{-11}$~n/g/s/ppb. 
\label{table:results}}
\begin{footnotesize}
\begin{tabular}{|c|c|c|cccc|}
\hline
Material & Formula / & Code & \multicolumn {4}{c|} {Neutron yield, n/g/s/ppb} \\
      & composition & & U & U$_{early}$ & U$_{late}$ & Th \\ \hline
\multirow{4}{*}{PTFE} & \multirow{4}{*}{C$_2$F$_4$} & EMPIRE3.2.3 & $9.39\times10^{-10}$ & $1.49\times10^{-10}$ & $7.90\times10^{-10}$ & $3.77\times10^{-10}$ \\
& & EMPIRE3.2.3 + Exp & $9.68\times10^{-10}$ & $1.55\times10^{-10}$ & $8.13\times10^{-10}$ & $3.91\times10^{-10}$ \\
& & EMPIRE2.19 & $8.72\times10^{-10}$ & $1.36\times10^{-10}$ & $7.36\times10^{-10}$ & $3.50\times10^{-10}$ \\
& & TALYS1.9 & $10.2\times10^{-10}$ & $1.63\times10^{-10}$ & $8.58\times10^{-10}$ & $4.03\times10^{-10}$ \\ \hline

\multirow{4}{*}{Ceramics} & \multirow{4}{*}{Al$_2$O$_3$} & EMPIRE3.2.3 & $1.14\times10^{-10}$ & $1.12\times10^{-11}$ & $1.03\times10^{-10}$ & $5.45\times10^{-11}$ \\
& & EMPIRE3.2.3 + Exp & $1.36\times10^{-10}$ & $1.64\times10^{-11}$ & $1.19\times10^{-10}$ & $6.04\times10^{-11}$ \\
& & EMPIRE2.19 & $8.59\times10^{-11}$ & $7.76\times10^{-12}$ & $7.81\times10^{-11}$ & $4.32\times10^{-11}$ \\
& & TALYS1.9 & $9.48\times10^{-11}$ & $8.99\times10^{-12}$ & $8.58\times10^{-11}$ & $4.56\times10^{-11}$ \\ 
\hline

\multirow{4}{*}{Quartz} & \multirow{4}{*}{SiO$_2$} & EMPIRE3.2.3 & $2.07\times10^{-11}$ & $3.25\times10^{-12}$ & $1.75\times10^{-11}$ & $8.61\times10^{-12}$ \\
& & EMPIRE3.2.3 + Exp & $1.35\times10^{-11}$ & $1.58\times10^{-12}$ & $1.19\times10^{-11}$ & $6.21\times10^{-12}$ \\
& & EMPIRE2.19 & $1.59\times10^{-11}$ & $2.01\times10^{-12}$ & $1.39\times10^{-11}$ & $7.03\times10^{-12}$ \\
& & TALYS1.9 & $1.54\times10^{-11}$ & $2.01\times10^{-12}$ & $1.34\times10^{-11}$ & $6.75\times10^{-12}$ \\ \hline

\multirow{4}{*}{Titanium} & \multirow{4}{*}{Ti(100)} & EMPIRE3.2.3 & $3.39\times10^{-11}$ & $2.55\times10^{-12}$ & $3.13\times10^{-11}$ & $2.48\times10^{-11}$ \\
& & EMPIRE3.2.3 + Exp & $3.39\times10^{-11}$ & $2.52\times10^{-12}$ & $3.14\times10^{-11}$ & $2.46\times10^{-12}$ \\
& & EMPIRE2.19 & $2.55\times10^{-11}$ & $1.11\times10^{-12}$ & $2.44\times10^{-11}$ & $2.15\times10^{-11}$ \\
& & TALYS1.9 & $2.80\times10^{-11}$ & $1.21\times10^{-12}$ & $2.68\times10^{-11}$ & $2.33\times10^{-12}$ \\ \hline

\multirow{4}{*}{Copper} & \multirow{4}{*}{Cu(100)} & EMPIRE3.2.3 & $6.55\times10^{-13}$ & $2.25\times10^{-14}$ & $6.33\times10^{-13}$ & $9.16\times10^{-13}$ \\
& & EMPIRE3.2.3 + Exp & $3.64\times10^{-13}$ & $8.48\times10^{-15}$ & $3.56\times10^{-13}$ & $7.55\times10^{-13}$ \\
& & EMPIRE2.19 & $3.11\times10^{-13}$ & $8.42\times10^{-15}$ & $3.03\times10^{-13}$ & $9.70\times10^{-13}$ \\
& & TALYS1.9 & $3.89\times10^{-13}$ & $8.73\times10^{-15}$ & $3.80\times10^{-13}$ & $1.51\times10^{-12}$ \\ \hline

\multirow{4}{*}{Acrylic} & \multirow{4}{*}{C$_5$H$_8$O$_2$} & EMPIRE3.2.3 & $2.28\times10^{-11}$ & $5.18\times10^{-12}$ & $1.76\times10^{-11}$ & $7.71\times10^{-12}$ \\
& & EMPIRE3.2.3 + Exp & $1.18\times10^{-11}$ & $1.95\times10^{-12}$ & $9.84\times10^{-12}$ & $4.66\times10^{-12}$ \\
& & EMPIRE2.19 & $1.33\times10^{-11}$ & $2.46\times10^{-12}$ & $1.09\times10^{-11}$ & $5.12\times10^{-12}$ \\
& & TALYS1.9 & $1.78\times10^{-11}$ & $4.07\times10^{-12}$ & $1.37\times10^{-11}$ & $6.11\times10^{-12}$ \\ \hline

\multirow{4}{*}{Polyethylene} & \multirow{4}{*}{CH$_2$} & EMPIRE3.2.3 & $2.46\times10^{-11}$ & $5.60\times10^{-12}$ & $1.90\times10^{-11}$ & $8.30\times10^{-12}$ \\
& & EMPIRE3.2.3 + Exp & $1.22\times10^{-11}$ & $1.85\times10^{-12}$ & $1.03\times10^{-11}$ & $4.98\times10^{-12}$ \\
& & EMPIRE2.19 & $1.47\times10^{-11}$ & $2.72\times10^{-12}$ & $1.20\times10^{-11}$ & $5.69\times10^{-12}$ \\
& & TALYS1.9 & $2.00\times10^{-11}$ & $4.64\times10^{-12}$ & $1.54\times10^{-11}$ & $6.83\times10^{-12}$ \\ \hline

\multirow{2}{*}{Stainless} & Fe(66),Cr(17) & EMPIRE3.2.3 & $7.18\times10^{-12}$ & $4.38\times10^{-13}$ & $6.74\times10^{-12}$ & $5.78\times10^{-12}$ \\
& Ni(12),Mn(2) & EMPIRE3.2.3 + Exp & $7.05\times10^{-12}$ & $4.21\times10^{-13}$ & $6.63\times10^{-12}$ & $5.69\times10^{-12}$ \\
\multirow{2}{*}{steel} & Mo(2),Si(1) & EMPIRE2.19 & $5.30\times10^{-12}$ & $2.13\times10^{-13}$ & $5.09\times10^{-12}$ & $6.05\times10^{-12}$ \\
& & TALYS1.9 & $6.29\times10^{-12}$ & $2.33\times10^{-13}$ & $6.06\times10^{-12}$ & $7.91\times10^{-12}$ \\ \hline

\end{tabular}
\end{footnotesize}
\end{center}
\end{table}

For most materials studied here, the 'new' cross-sections, either from EMPIRE3.2.3 or TALYS1.9 result in higher neutron yields compared to the 'old' cross-sections from EMPIRE2.19. When replacing EMPIRE3.2.3 calculations with experimental data at low energies (where available), the results largely depend on which data set is used. The spread of data is very large (larger than for calculated cross-sections in realistic models) and there is no obvious choice of the data set to use. Measurements of the cross-sections are usually limited to the total cross-sections and do not provide transition probabilities to various excited states (not to high excited states anyway) so the use of a model is unavoidable to obtain the correct neutron spectrum. Hence the results for 'EMPIRE3.2.3 + Exp' are shown here for illustration only using one of the data sets for a particular isotope. (One possible exception to this is the cross-section on $^{13}$C that has been measured to high precision up to highest alpha energies, as shown in Figure~\ref{fig:cs-models-carbon}. This can be combined with transition probabilities from a model.)

Although there is no specific tendency when comparing EMPIRE3.2.3 and TALYS cross-sections for different isotopes, for most materials selected for Table \ref{table:results} EMPIRE3.2.3 gives higher neutron yields than TALYS1.9. There is no obvious reason for choosing specific model: EMPIRE3.2.3 or TALYS1.9. Both codes are evolving and the difference between particular code versions may exceed the difference between the two different codes. For conservative estimates of neutron background, the cross-sections from EMPIRE3.2.3 can be used. For most materials the difference between EMPIRE3.2.3 and TALYS1.9 does not exceed 20\% but for quartz it is up to 30\% for uranium decay chain.

\section{Conclusions}
\label{sec-conclusions}

Cross-sections of ($\alpha,n$) reactions from different codes and models have been investigated. The study has covered several models in the EMPIRE3.2.3 code, the old EMPIRE2.19, the recent version of TALYS1.9 and experimental data. Experimental data show a large spread, even larger than that for different models, and do not allow to select the most reliable model to use. The cross-sections calculated with the most recent versions of EMPIRE3.2.3 (with parameters recommended by the code authors) and TALYS1.9 (with default parameters) may differ by about 20\% with similar differences in transition probabilities to different final states. 

Neutron yields for different materials have been calculated with different sets of cross-sections using modified SOURCES4A code and found to follow the behaviour of the cross-sections. Most recent versions of the cross-section codes EMPIRE3.2.3 and TALYS1.9 give higher neutron yields that the old EMPIRE2.19. The difference between EMPIRE3.2.3 and TALYS1.9 does not exceed 20\% for most materials with the former usually giving higher neutron yields (for materials presented in Table \ref{table:results}).

\section{Acknowledgments}

The authors would like to thank Science and Technology Facilities Council (STFC, UK) and the University of Sheffield for financial support.


\bibliographystyle{elsarticle-num}
\bibliography{neutrons-alphan}{}%

\end{document}